\documentclass[
reprint,
superscriptaddress,
amsmath,amssymb,
aps,
prl,
]{revtex4-2}

\usepackage{graphicx,import}
\usepackage{dcolumn}% Align table columns on decimal point
\usepackage{upgreek}
\usepackage{physics}
\usepackage{amsmath}
\usepackage{setspace}%Ê¹ÓÃ¼ä¾àºê°ü
\usepackage{bm}
\usepackage{xcolor}
\usepackage{ulem}
\usepackage{lipsum}

\usepackage[breaklinks=true,colorlinks=true,linkcolor=blue,urlcolor=blue,citecolor=blue]{hyperref}
\allowdisplaybreaks[4]
\UseRawInputEncoding

\begin{document}

\preprint{APS/123-QED}

\title{Subtraction and Addition of Propagating Photons by Two-Level Emitters}

\author{Mads M. Lund}
\thanks{These authors contributed equally to this work. \\ $^\dagger$Contact author: fanyangphys@gmail.com}
\affiliation{Center for Complex Quantum Systems, Department of Physics and Astronomy, Aarhus University, DK-8000 Aarhus C, Denmark}

\author{Fan Yang$^\dagger$}
\thanks{These authors contributed equally to this work. \\ $^\dagger$Contact author: fanyangphys@gmail.com}
%\affiliation{Contact author: fanyangphys@gmail.com}
\affiliation{Niels Bohr Institute, University of Copenhagen, Blegdamsvej 17, 2100 Copenhagen, Denmark}

\author{Victor Rueskov Christiansen}
\affiliation{Center for Complex Quantum Systems, Department of Physics and Astronomy, Aarhus University, DK-8000 Aarhus C, Denmark}

\author{Danil Kornovan}
\affiliation{Center for Complex Quantum Systems, Department of Physics and Astronomy, Aarhus University, DK-8000 Aarhus C, Denmark}

\author{Klaus M{\o}lmer}
%\email{klaus.moelmer@nbi.ku.dk}
\affiliation{Niels Bohr Institute, University of Copenhagen, Blegdamsvej 17, 2100 Copenhagen, Denmark}

	\newcommand{\hatb}{\hat{b}}
	\newcommand{\hata}{\hat{a}}
	\newcommand{\hatsigma}{\hat{\sigma}}
    \newcommand{\hatc}{\hat{c}}
	\newcommand{\hatau}{\hat{a}_u}

	%\date{\today}
	
\begin{abstract}
Coherent manipulation of quantum states of light is key to photonic quantum information processing. In this Letter, we show that a passive two-level nonlinearity suffices to implement non-Gaussian quantum operations on propagating field modes. In particular, the collective light-matter interaction can efficiently extract a single photon from a multi-photon input wave packet to an orthogonal temporal mode. We accurately describe the single-photon subtraction process by elements of an intuitive quantum-trajectory model. By employing this process, quantum information protocols gain orders of magnitude improved efficiency over heralded schemes with linear optics. The reverse process can be used to add photons one-by-one to a single wave-packet mode and compose arbitrarily large Fock states with a finite total success probability $>96.7\%$.
\end{abstract}

\maketitle

{\it Introduction}.---Propagating photons are ideal carriers of quantum information, since they can be precisely manipulated, detected, and distributed in a scalable manner \cite{flamini2018photonic,slussarenko2019photonic,PhysRevX.5.041017,wang2020integrated}. While Gaussian operations such as beam splitting \cite{zhong2020quantum} and squeezing \cite{andersen201630} are well established, efficient non-Gaussian operations are still under active  development, as they are essential to exhaust the potential of bosonic fields in quantum computing and simulation \cite{PhysRevLett.88.097904,PhysRevLett.89.137903,PhysRevLett.89.137904,PhysRevA.66.032316,PhysRevLett.109.230503}.

Single-photon subtraction and addition are such non-Gaussian processes that have received considerable attention \cite{nphys1199,nphoton.2010.1,PhysRevA.90.013821,PhysRevX.7.031012,ra2020non}. Based on beam splitters, squeezed light sources and photon detection, heralded schemes exist for this purpose \cite{zavatta2004,ourjoumtsev2006generating,PhysRevLett.97.083604,PhysRevLett.98.030502,PhysRevA.80.053822}, but they succeed only with low probabilities. A number of nonlinear optical setups have been proposed to achieve more favorable deterministic operation, including so-called active schemes where an input pulse may be converted to a single cavity mode, affected by a subsequent unitary cavity QED interaction \cite{zou2006simple,PhysRevA.86.032311,PhysRevLett.110.210504}. Passive schemes \cite{PhysRevA.88.033832,Du:20,Rosenblum2016,PhysRevLett.117.223001,stiesdal2021,PhysRevLett.120.113601}, in contrast may operate on the input pulse in a more robust, autonomous manner. However, the interaction of input photons with the nonlinear medium over time usually populates multiple field modes and results in a reduced purity of the states generated \cite{PhysRevLett.110.153601,PhysRevLett.125.143601}, which severely limits their practical applications in quantum information processing.

\begin{figure}[b]
    \centering
    \includegraphics[width=\linewidth]{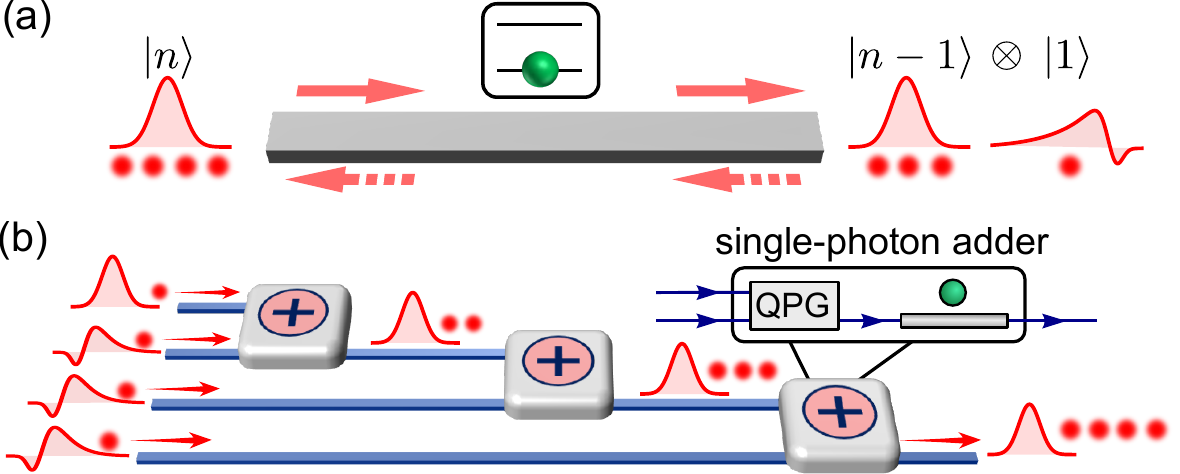}
    \caption{(a) Schematic of the single-photon subtraction and addition mediated by a TLE. (b) Construction of a high photon number state from individual photons. The single-photon adder is composed of a quantum pulse gate (QPG) and a TLE.}
    \label{fig:fig1}
\end{figure}

The saturation-type nonlinearity of a two-level emitter (TLE) coupled to a unidirectional continuum field \cite{RevModPhys.89.021001,RevModPhys.90.031002,RevModPhys.95.015002} is conjectured to support single-photon subtractions, based on the intuitive argument, that the emitter can only absorb a single photon at a time. In this Letter, we demonstrate that a passive two-level nonlinearity is, indeed, sufficient for single-photon subtraction. While all photons taking part in the dynamics could potentially scatter into a vast number of modes, we show that, provided the optimal duration of the input pulse, a single photon is converted to a temporal mode orthogonal to the original mode still occupied by the remaining photons [Fig.~\ref{fig:fig1}(a)]. This makes the conjugate process, adding a single photon to a Fock-state pulse by the same component [dashed arrows in Fig.~\ref{fig:fig1}(a)], equally efficient.

\begin{figure*}
    \centering
    \includegraphics[width=\textwidth]{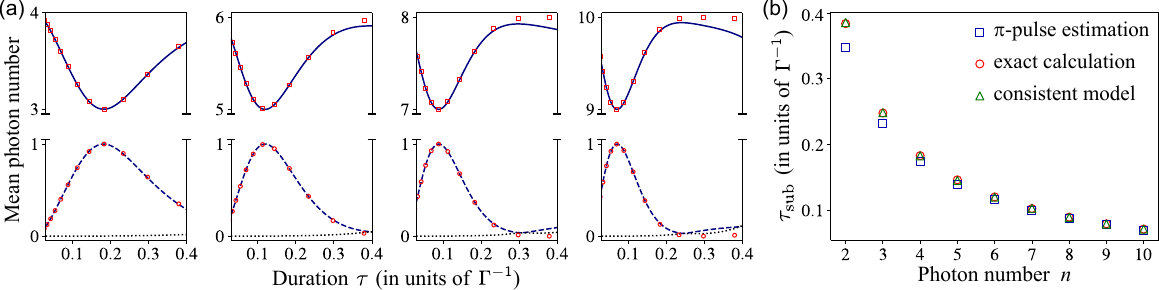}
    \caption{Scattering of a single-mode input pulse by a TLE. (a) Mean photon number in different output modes as a function of the duration $\tau$ of the input Gaussian pulse. The four panels (from left to right) show the results for input photon number $n=4$, $6$, $8$ and $10$, respectively. The black solid line, the black dashed line, and the black dotted line show the mean photon number in the dominant mode ($\bar{n}_1$), the second dominant mode ($\bar{n}_2$), and the residual modes ($\sum_{i>2}\bar{n}_i$), respectively. The red squares and circles are obtained from a consistent model, presented in the text. (b) Duration $\tau_\mathrm{sub}$ for achieving the optimal single-photon subtraction. The blue squares correspond to a simple $\pi$-pulse assumption. The red circles are obtained from the exact calculation of $\rho_{n-1,n-1}^a$. The green triangles are obtained from our consistent model.}
    \label{fig:fig2}
\end{figure*}
{\it Single-photon subtraction}.---We consider the scattering of a unidirectional continuous field $\hat{\mathcal{E}}(t)$ by a two-level emitter (TLE), as can be visualized by the chiral waveguide QED configuration shown in Fig.~\ref{fig:fig1}(a). The incoming field is in an $n$-photon Fock state $\ket{n}=(\hata^\dagger)^n \ket{0}/\sqrt{n!}$, where $\hata^\dagger = \int dt~\phi_a(t) \hat{\mathcal{E}}^\dagger(t)$ creates a single photon in a given temporal mode $\phi_a(t)$. To investigate the scattering dynamics of such an input field pulse, we employ an adaption of the input-output formalism \cite{qpm1,qpm2,victor_interaction}, which allows calculation of the correlations in the output field $\hat{\mathcal{E}}_\mathrm{out}(t)$. In general, the output photons will not be restricted to the input mode but form a complicated time-frequency entangled state. To analyze the multimode character of the corresponding state, we carry out the Karhunen-Lo\`eve expansion of the first order correlation function, $\langle \hat{\mathcal{E}}^\dagger_\mathrm{out}(t)\hat{\mathcal{E}}_\mathrm{out}(t^\prime)\rangle=\sum_i \bar{n}_i\phi_i^*(t)\phi_i(t^\prime) $, where $\phi_i(t)$ denotes a set of orthonormal temporal modes and $\bar{n}_i$ denotes the mean photon number in each mode \cite{perf2psplit}.

Figure \ref{fig:fig2}(a) shows the results of this expansion for different input Fock states $n=$ 4, 6, 8, 10, where the mean photon number in the most populated mode (upper solid lines), the second most populated mode (middle dashed lines), and the rest of the modes (lower dotted lines) is plotted as a function of the duration $\tau$ of the input Gaussian pulse $\phi_a(t)\propto e^{-(t-4\tau)^2/2\tau^2}$. Evidently, for pulse durations shorter than the lifetime ($\Gamma^{-1}$) of the TLE, the first two modes ($i=1,2$) dominate the photon population in the output field with $\bar{n}_1+\bar{n}_2\approx n$. More interestingly, $\bar{n}_1$ and $\bar{n}_2$ attain values $\bar{n}_1\approx n-1$ and $\bar{n}_2\approx 1$ at specific durations $\tau_\mathrm{sub}$, where the mode function $\phi_1(t)$ also approaches the input mode $\phi_a(t)$. Such a photon-number splitting is indicative of a perfect single-photon subtraction from the input field. To confirm that a single quantum has been removed, we examine the reduced density matrix $\hat{\rho}^{a}$ for output photons residing in the input mode $\phi_a(t)$ \cite{qpm1,qpm2,victor_interaction}. We find that the element $\rho_{n-1,n-1}^a=\langle n-1|\hat{\rho}^{a}|n-1\rangle$ indeed approaches unity ($>0.996$) at the optimal duration $\tau_\mathrm{sub}$, suggesting a successful subtraction of one photon from the input Fock state. Here, the subtracted photon is converted into a temporal mode $\phi_b(t)=\phi_2(t)$ orthogonal to the input mode $\phi_a(t)$. The output state is thus given by $\ket{n-1}\otimes\ket{1}=(\hata^\dagger)^{n-1} \hatb^\dagger\ket{0}/\sqrt{(n-1)!}$, where $\hatb^\dagger = \int dt~\phi_b(t) \hat{\mathcal{E}}^\dagger(t)$ creates a single photon in $\phi_b(t)$.

\begin{figure}
    \centering
    \includegraphics[width=\linewidth]{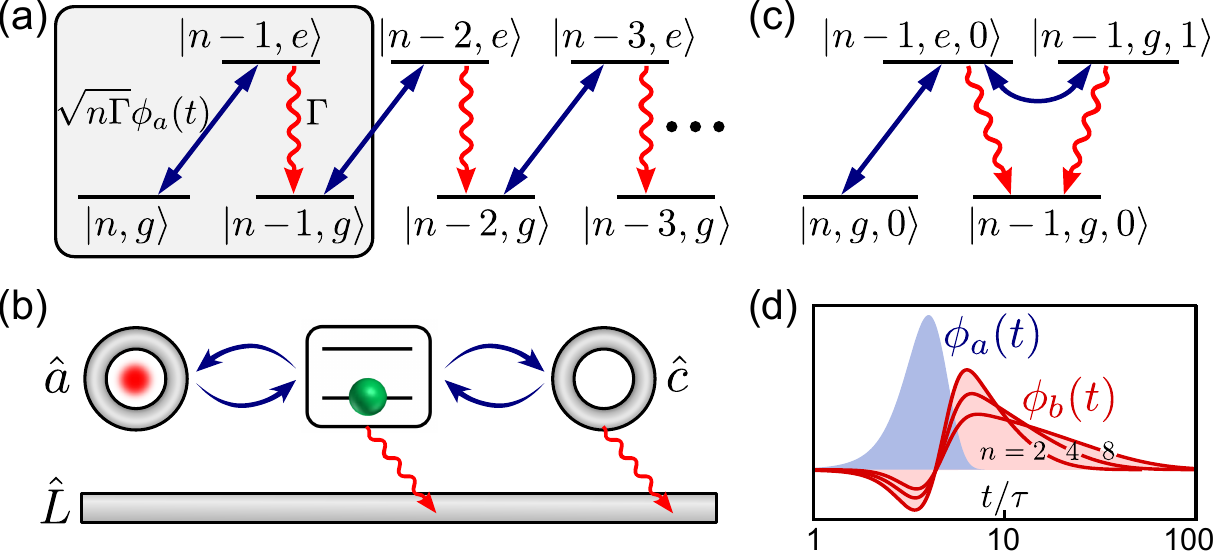}
    \caption{(a) Level scheme for the intuitive JC model, where the shaded block indicates the truncated subspace populated by a short input pulse. (b) and (c) illustrate our consistent model incorporating an auxiliary mode and its (truncated) level scheme, respectively. (d) Shape of the input and output mode $\phi_a(t)$ and the singly occupied output mode $\phi_b(t)$ for the input photon number $n=2,4,8$. The linewidth $\Gamma$ is chosen to achieve the optimal subtraction for each $n$, and the horizontal axis is plotted in logarithmic scale.}
    \label{fig:fig3}
\end{figure}

To understand why the TLE can behave as a perfect single-photon subtractor and how the optimal duration scales with the input photon number, we first consider a simplistic, intuitive model. By simply replacing the continuum with a single mode $\hat{\mathcal{E}}(t)\approx \phi_a(t)\hat{a}$, the light-matter interaction is described by a Jaynes-Cummings (JC) Hamiltonian $\hat{H} = i\sqrt{\Gamma} \phi_a(t)(\hata^\dagger\hatsigma_- - \hatsigma_+\hata)$, where $\hatsigma_\pm$ denotes the spin raising and lowering operator of the TLE. Assuming an independent spontaneous emission process, the scattering can be depicted by the level diagram shown in Fig.~\ref{fig:fig3}(a), where the straight and the wavy arrows represent the coherent and the incoherent part of the dynamics, respectively. The interaction between the input field and the TLE first converts a photon from the single pulse mode oscillator into an atomic excitation $\ket{n,g}\rightarrow \ket{n-1,e}$. Then the spontaneous emission removes the TLE excitation from the system and leads to $\ket{n-1,g}$, which may be subsequently re-excited to $\ket{n-2,e}$ by the tail of the input field. For a short duration $\tau\lesssim\Gamma^{-1}$, the input pulse traverses the TLE before the completion of the spontaneous emission, such that only the first sector of the ladder [shaded block in Fig.~\ref{fig:fig3}(a)] is relevant. In this simplified picture, single-photon subtraction will therefore occur if the quantum Rabi oscillation dynamics executes a $\pi$-pulse [$2\int dt\sqrt{n\Gamma}\phi_a(t)=\pi$] that perfectly transfers the initial state into $\ket{n-1,e}$. For Gaussian input modes, this yields an optimal duration $\tau_\pi=\pi^{3/2}/(8\Gamma n)$.

While this intuitive model provides a qualitative account of the observed photon subtraction, as shown in Fig.~\ref{fig:fig2}(b), it fails to reproduce the precise optimal duration $\tau_\mathrm{sub}$ for small and intermediate photon numbers. More crucially, the model does not account properly for the spatiotemporal behavior of the subtracted photon and ensure that it is orthogonal to the input mode. These omissions stem from the single-mode approximation in the simplistic JC model, as the propagating light field is inherently multimode.

We develop here a consistent theory which can yield quantitative predictions while preserving the intuition offered by the JC model. To properly account for the interaction of the TLE with the multimode continuous field, we employ a recently developed approach \cite{victor_interaction}, which corroborates the validity of the intuitive single-mode JC model, giving rise, however, to the additional coupling to an auxiliary single mode $\hat{c}$. In the interaction picture, the system dynamics is described by a Lindblad master equation with a Hamiltonian 
\begin{equation}
\hat{H}=i\sqrt{\Gamma}\phi_a^*(t)\left[\hat{a}^\dagger\hat{\sigma}_- + \cot2\theta(t)\hat{c}^\dagger\hat{\sigma}_-\right]+\mathrm{H.c.},\label{eq:H_int}
\end{equation}
and a single Lindblad dissipation term  
\begin{align}
    \hat{L} = \sqrt{\Gamma}\hatsigma_- - 2 \phi_a(t)\csc 2\theta (t) \hat{c},\label{eq:L_c}
\end{align}
where $\theta(t)$ is defined as $\sin^2\theta(t)=\int_{0}^t d\xi |\phi_a(\xi)|^2$. The role of the auxiliary mode is illustrated in Fig.~\ref{fig:fig3}(b): it interacts coherently with the TLE, and its leakage interferes with the spontaneous emission from the emitter via the collective decay $\hat{L}$. This interference accurately describes the photon loss from the input mode, i.e., the subtracted photon and its wavefunction. Inclusion of the auxiliary mode results in a more complicated ladder of quantum states, but similar to Fig.~\ref{fig:fig3}(a), a truncation can be made for short input pulses. The truncated subspace is displayed in Fig.~\ref{fig:fig3}(c), where the initial state $\ket{n,g,0}$ with the auxiliary mode in the vacuum state evolves via intermediate states $\ket{n-1,e,0}$ and $\ket{n-1,g,1}$ towards the photon-subtracted target state $\ket{n-1,g,0}$.

 We can now apply an efficient quantum-trajectory description of the dynamics \cite{Caneva_2015,zhang2018quantum,PhysRevResearch.4.023002,li2022control}, where the state follows a non-unitary evolution until the probabilistic occurrence of a quantum jump by $\hat{L}$ which removes an excitation from the system. As the truncated subspace allows only a single jump, the unnormalized wavefunction for the bath excitation (subtracted photon) is determined by
\begin{align}
    \Tilde{\phi}_b(t) = \bra{n-1,g,0}{\hat{L}\cdot\hat{U}_\mathrm{eff}(t,0)}\ket{n,g,0}.
    \label{eq:psitilde1}
\end{align}
Here, the non-unitary evolution operator $\hat{U}_\mathrm{eff}(t,0)$ is governed by $i\partial_t\hat{U}_\mathrm{eff}(t,0) = \hat{H}_\mathrm{eff}(t)\hat{U}_\mathrm{eff}(t,0)$, with the non-Hermitian Hamiltonian $\hat{H}_\mathrm{eff}=\hat{H}-i\hat{L}^\dagger\hat{L}/2$:
\begin{align}
    \hat{H}_\mathrm{eff}= i\begin{bmatrix}
        0 & \sqrt{n\Gamma}\phi_a^*(t) & 0\\
        -\sqrt{n\Gamma}\phi_a(t) & -\Gamma/2 & \sqrt{\Gamma}\phi_a(t)\tan\theta \\
        0 & \sqrt{\Gamma}\phi_a^*(t)\cot\theta & -2|\phi_a(t)\csc 2\theta|^2
    \end{bmatrix}.\nonumber
\end{align}
The evolved no-jump state $|\psi(t)\rangle=\hat{U}_\mathrm{eff}(t,0)\ket{n,g,0}=|\psi(t)\rangle = C_1(t)|n,g,0\rangle+C_2(t)|n-1,e,0\rangle+C_3(t)|n-1,g,1\rangle$ ultimately attains $C_1(\infty)|n,g,0\rangle$, from which we can construct the output state of the photonic field  
\begin{align}
\ket{\Psi_\mathrm{out}} = C_1(\infty)\frac{(\hata^\dagger)^n}{\sqrt{n!}}\ket{0}+\sqrt{P}\frac{(\hata^\dagger)^{n-1}}{\sqrt{(n-1)!}}\hatb^\dagger\ket{0},
    \label{eq:psiout}
\end{align}
where $\hatb^\dagger$ creates a photon in the new temporal mode fed by the dissipation term $\hat{L}$ (not to be confused with the $C_3$ auxiliary photon component), and $P = 1-|C_1(\infty)|^2=\int dt\, |\Tilde{\phi}_b(t)|^2$ is the subtraction efficiency with $\Tilde{\phi}_b(t) = \sqrt{\Gamma}C_2(t) -2\phi_a(t)\csc 2\theta(t)C_3(t)$ obtained from Eq.~\eqref{eq:psitilde1}. Compensating for the dispersion caused by the auxiliary mode \cite{supply}, the normalized wavefunction for mode $\hatb$ reads
\begin{align}
    \phi_b (t) = \frac{1}{\sqrt{P}}\left[\Tilde{\phi}_b(t)-\phi_a(t)\int_t^\infty d\xi\, \phi_a^*(\xi)\Tilde{\phi}_b(\xi)\csc^2\theta(\xi)\right].\nonumber
\end{align}
As illustrated in Fig.~\ref{fig:fig3}(d), while $\phi_b(t)$ has an exponential tail $\sim e^{-\Gamma t/2}$ being fed by the TLE spontaneous emission, its shape at earlier times is significantly modified and secures the orthogonality condition $\int dt\, \phi_a^*(t)\phi_b(t)=0$.

The quantitative performance of our consistent theory is tested in Fig.~\ref{fig:fig2}(a), where we use Eq.~\eqref{eq:psiout} to obtain the output field correlation function and determine the two eigenmodes. The results (indicated by the squares and circles) agree well with the exact mode decomposition, while a visible deviation only appears at a large duration $\tau$. In this region, the state $\ket{n-1,g,0}$ accumulates non-negligible populations while the TLE still interacts with the input mode, which drives the system out of the truncated subspace in Fig.~\ref{fig:fig3}(c), and causes multiphoton and multimode subtracted components. We are interested in the single-photon subtraction regime where our consistent model can predict with high precision the optimal duration $\tau_\mathrm{sub}$ of the input mode [Fig.~\ref{fig:fig2}(c)] as well as the shape of the output mode $\hatb$ [Fig.~\ref{fig:fig3}(d)]. In the Supplemental Material \cite{supply}, we extend the quantum-trajectory formalism beyond the truncated subspace considered here to accurately describe the photon subtraction also for long pulses with $\tau>\Gamma^{-1}$.

\begin{figure}
    \centering
    \includegraphics[width=\linewidth]{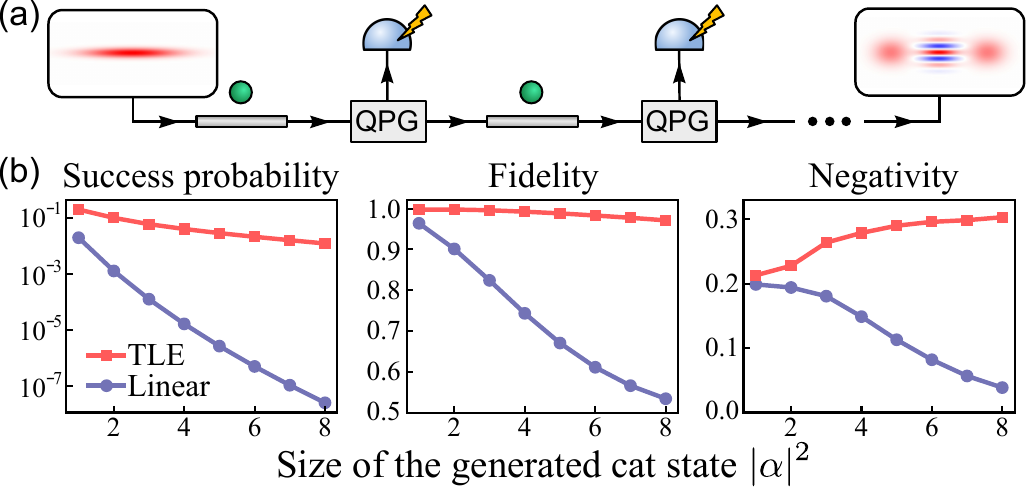}
    \caption{(a) Generation of cat states from a squeezed vacuum state by successive single-photon subtractions. (b) Performance of the TLE scheme ($\Gamma\tau=0.04$) and the linear scheme ($R=0.01$) for a 10~dB initial squeezing. The output state is optimally unsqueezed to achieve the largest fidelity. The input and output blocks in (a) display the Wigner functions for the input state and the generated cat state ($|\alpha|^2$=6).}
    \label{fig:fig4}
\end{figure}
{\it Efficient generation of non-Gaussian states}.---In the literature, the photon subtraction processes described by an annihilation operator $\hata$ is a highly non-Gaussian operation, which can be applied on a wide class of input states to distill states with a non-positive Wigner function \cite{ourjoumtsev2006generating,PhysRevLett.97.083604,PhysRevLett.98.030502}. In linear optics, $\hat{a}$ is implemented as the effect of a (rare) heralding process in which a single photon is detected in the signal reflected by a beam splitter. To suppress multiphoton subtraction events, one has to choose a relatively small reflectance, which inevitably reduces the success probability. In contrast, multiphoton subtraction is intrinsically suppressed by the two-level nonlinearity in the TLE-based subtraction discussed above. Our scheme thus holds promise to enable more efficient schemes for heralded production of non-classical states.

The TLE-based photon subtraction transforms a superposition state $\sum_n c_n|n\rangle$ into $\sum_n c_n f(n)|n-1\rangle$, conditioned on successful subtraction of a single photon in a given temporal mode \cite{supply}. When $n\Gamma\tau < 1$, the filtering function $f(n)$ approaches the solution of the quantum Rabi oscillations, and hence, the ideal one of a photon annhilation, i.e., $f(n)\approx \sin\sqrt{2\pi^{1/2} n\Gamma\tau}\sim\sqrt{n}$. To demonstrate the potential for generation of non-classical states in this interaction regime, we consider generating a Schr\"odinger cat state $|\mathrm{cat}\rangle\propto [|\alpha\rangle+(-1)^M|-\alpha\rangle]$ from a squeezed vacuum state via $M$ successive photon subtractions \cite{PhysRevA.103.013710}, where $|\alpha\rangle$ denotes a coherent state with mean photon number $|\alpha|^2=M$. The setup is illustrated in Fig.~\ref{fig:fig4}(a), where a quantum pulse gate (QPG)  \cite{eckstein2011quantum,silberhorn1,silberhorn2} is used to distill the state in the input temporal mode $\hata$, while detection of photons in orthogonal temporal modes heralds a successful photon subtraction. The relation between the heralded state $\hat{\rho}_\mathrm{out}$ and the input state $\hat{\rho}_\mathrm{in}$ for each subtraction process is given by $\hat{\rho}_\mathrm{out}=\hat{\rho}^a-\hat{U}_\mathrm{eff}(\infty,0)\hat{\rho}_\mathrm{in}\hat{U}_\mathrm{eff}^\dagger(\infty,0)$, where $\hat{\rho}^a$ follows the master equation evolution [Eqs.~\eqref{eq:H_int} and \eqref{eq:L_c}], while the second term describes the evolution of $\hat{\rho}_\mathrm{in}$ conditioned on zero detector clicks \cite{supply}.

We compare the typical performance of our scheme and the linear scheme in Fig.~\ref{fig:fig4}(b), where the parameters ($\Gamma \tau = 0.04$ and reflectance $R=0.01$) are chosen after balancing the trade-off between efficiency and operation fidelity. As expected, the TLE-based subtractor has a significantly larger success probability as well as a better scaling with the number of operations. In addition, the fidelity and negativity of the generated state remain high in the TLE-based scheme, as multiphoton subtraction events are largely suppressed. The state fidelity can be further increased at the cost of a reduced success probability by merely choosing a smaller $\Gamma\tau$. Without further optimization, the performance of the TLE-based subtraction is comparable to an advanced, generalized subtraction scheme \cite{PhysRevA.103.013710}, which needs number-resolved photon detectors that are not required here.

\begin{figure}[b]
    \centering
\includegraphics[width=\linewidth]{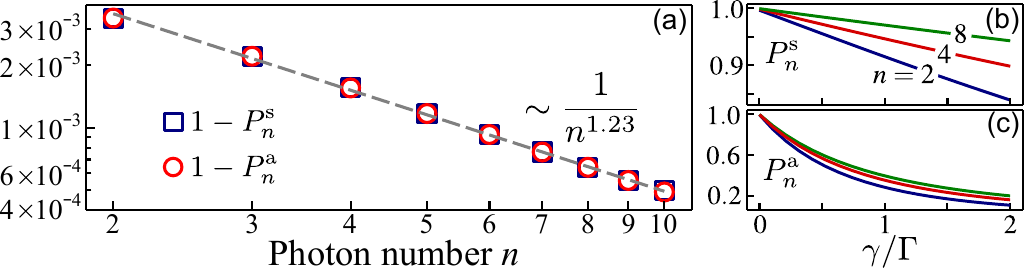}
    \caption{(a) Scaling of the failure probability for optimal single-photon subtractions (blue squares) and additions (red squares). The dashed line denotes the power-law fitting to the data. (b) and (c) show the success probability of optimal single-photon subtractions and additions as a function of the imperfect coupling strength $\gamma$.}
    \label{fig:fig5}
\end{figure}

{\it Deterministic photon addition}.---The parity-time symmetry of the system allows us to apply a conjugate operation of the photon subtraction, i.e., deterministically adding a single photon in mode $\phi_b^{T}(t) = \phi_b^*(T-t)$ to mode $\phi_a^{T}(t) = \phi_a^*(T-t)$ carrying $n-1$ photons [see dashed arrows in Fig.~\ref{fig:fig1}(a)], where $T$ should be sufficiently large to complete the pulses. The success probability $P_n^\mathrm{a}=\rho_{n,n}^{a}$ of the single-photon addition is identical to the success probability $P_n^\mathrm{s}=\rho_{n-1,n-1}^{a}$ of the single-photon subtraction due to a generalized reciprocity theorem \cite{supply}. This is verified by the numerical simulation shown in Fig.~\ref{fig:fig5}, where the success probability $P_n^\mathrm{a}$ and $P_n^\mathrm{s}$ at the optimal duration coincide and approach unity as $n$ increases. Here, we identify a power-law scaling of the failure probability $1-P_n^\mathrm{a}\sim {n^{-\beta}}$ with $\beta\approx 1.23$. We note that the exponent $\beta>1$ is formally due to the interaction with the auxiliary mode, without which we would expect $\beta=1$ from the intuitive JC model.

The large success probability and its superior scaling property allows one to efficiently compose a large Fock state from individual single photons. Specifically, by cascading the single-photon addition $|n-1\rangle\otimes|1\rangle\rightarrow|n\rangle\otimes|0\rangle$ from $n=2$ to $n=M$, one can create an $M$-photon Fock state with a success probability $P_{M} = \prod_{n=2}^{M}P_n^\mathrm{a}$. The scaling factor $\beta>1$ obtained in Fig.~\ref{fig:fig5}(a) implies that ideally an arbitrarily large Fock state can be generated with a finite success probability: for all $M$, $P_{M}>96.7\%$, in stark contrast to exponentially decreasing success probabilities in a probabilistic scheme. The practical implementation of our scheme is illustrated in Fig.~\ref{fig:fig1}(b), where high-purity single photons in different pulse shapes are independently generated \cite{lodahl2,spsc1,spsc2,spsc3} and subsequently combined by a QPG, followed by the scattering on the TLE with an optimal interaction strength $\Gamma\propto 1/n$.

So far, the discussion is based on an ideal model where the TLE perfectly couples to the forward propagating mode. For an imperfect coupling, the scattering into the backward mode or free-space modes can be described by an additional decay rate $\gamma$. Figures~\ref{fig:fig5}(b) and \ref{fig:fig5}(c) show the performance of the subtraction and addition with a finite $\gamma$. While both $P_n^\mathrm{s}$ and $P_n^\mathrm{a}$ decrease as $\gamma/\Gamma$ increases, they are no longer identical. In particular, the subtraction process appears to be much more robust against imperfections than the addition. This is because $P_n^\mathrm{s}$ does not depend on details of the field modes orthogonal to the input one. As $n$ increases further, $P_n^\mathrm{s}$ will approach unity due to suppressed multiphoton subtractions, while $P_n^\mathrm{a}$ will approach the branching ratio $\beta=\Gamma/(\gamma+\Gamma)$. In state-of-the-art waveguide QED setups, $\beta$ can be made up to $\sim99\%$ \cite{PhysRevB.100.035311}, which makes the implementation of the subtraction/addition scheme quite achievable.

{\it Conclusion and outlook}.---In summary, we have examined the scattering of a unidirectional wave packet of light by a two-level emitter, and identified a near-perfect photon subtraction and addition to occur for pulses in the right temporal mode. We have offered intuitive and consistent descriptions of the process, explaining its crude features and offering very accurate accounts of its origin. The deterministic feature of the demonstrated process enables efficient generation of large-size cat states and Fock states, offering indispensable resources for photonic quantum information processing \cite{Hastrup2022all,ourjoumtsev2007generation}. The results and analysis apply also for microwave and acoustic waves scattering on, e.g., superconducting elements, and can be generalized to treat systems containing multiple waveguide channels \cite{lindkvist2014scattering}. Our effective analysis with an auxiliary mode provides a bridge between cavity and waveguide QED and causes optimism for further exploration of quantum nonlinear optics with travelling pulses.

\begin{acknowledgments}
We thank Anton L. Andersen, \v{S}imon Br\"auer, Ying Wang, and Anders S. S{\o}rensen for helpful discussions. This work is supported by the Carlsberg Foundation through the ``Semper Ardens'' Research Project QCooL.
\end{acknowledgments}

\bibliography{main_text_bib}

\end{document}

\preprint{APS/123-QED}
	
\title{Supplementary Material for ``Subtraction and Addition of Propagating Photons by Two-Level Emitters''}

\author{Mads M. Lund}
\thanks{These authors contributed equally to this work.}
\affiliation{Center for Complex Quantum Systems, Department of Physics and Astronomy, Aarhus University, DK-8000 Aarhus C, Denmark}

\author{Fan Yang}
\thanks{These authors contributed equally to this work.}
\affiliation{Niels Bohr Institute, University of Copenhagen, Blegdamsvej 17, 2100 Copenhagen, Denmark}

\author{Victor Rueskov Christiansen}
\affiliation{Center for Complex Quantum Systems, Department of Physics and Astronomy, Aarhus University, DK-8000 Aarhus C, Denmark}

\author{Danil Kornovan}
\affiliation{Center for Complex Quantum Systems, Department of Physics and Astronomy, Aarhus University, DK-8000 Aarhus C, Denmark}

\author{Klaus M{\o}lmer}
%\email{klaus.moelmer@nbi.ku.dk}
\affiliation{Niels Bohr Institute, University of Copenhagen, Blegdamsvej 17, 2100 Copenhagen, Denmark}

\newcommand{\hatb}{\hat{b}}
\newcommand{\hata}{\hat{a}}
\newcommand{\hatsigma}{\hat{\sigma}}
\newcommand{\hatc}{\hat{c}}
\newcommand{\hatau}{\hat{a}_u}
\newcommand{\hatL}{\hat{L}}

	%\date{\today}
	
\begin{abstract}
This supplementary material provides detailed accounts of technical elements in the main text, including: derivation of the effective master equation (Sec.~I); a general quantum trajectory formalism (Sec.~II); approximate analytical result for a two-photon input state (Sec.~III); distillation of non-Gaussian states (Sec.~IV); proof of the reciprocity theorem (Sec.~V); and considerations for experimental realizations (Sec.~VI).
\end{abstract}

\maketitle
	
\section{Effective master equation}\label{sec:SecI}
\begin{figure}[b]
	\centering
	\includegraphics[width=\linewidth]{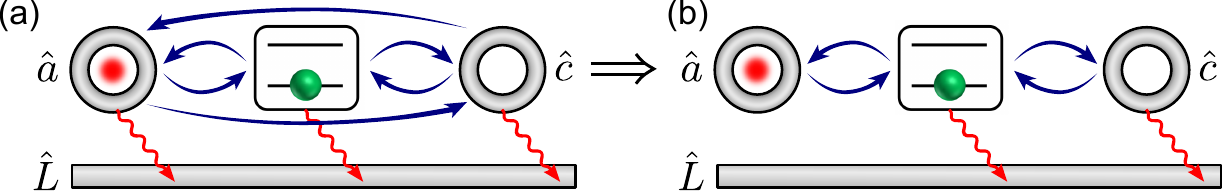}
	\caption{Depiction of the combined system of an input pulse interacting with a two-level system and subsequent retrieval of a specific mode in the scattered output field. The bent, blue arrows indicate the interactions in the Hamiltonian, while the curvy, red arrows indicate emission to the waveguide given by the Lindblad damping operators. (a) The combined system given by Eqs.~\eqref{eq:H} and \eqref{eq:L}. (b) The combined system in the interaction picture given by Eqs.~\eqref{eq:H-int} and \eqref{eq:L-int}.}
	\label{fig:fig1}
\end{figure}

In this section, we derive an effective master equation for the interaction between a travelling pulse of quantum radiation and a localized quantum system. As described in Refs.~\cite{qpm1, qpm2}, the interaction with a pulse can be efficiently modeled through an equivalent picture, where an upstream cavity leaks the  pulse of radiation on the system. The downstream scattered radiation can then be picked up by equivalent output cavities, each picking up a single mode of radiation (note that these cavities are virtual, theoretical constructions that allow determination of the quantum state contents of the actual travelling wavepacket modes in an experiment). To leak a desired incoming mode $\phi(t)$, the input virtual cavity requires a time-dependent coupling strength of
\begin{equation}
    g_\phi(t) = \frac{\phi^*(t)}{\sqrt{1 - \int_0^t dt' |\phi(t')|^2}},
\end{equation}
while the capture of a given scattered output mode $\psi(t)$ requires an output virtual cavity with time-dependent coupling strength given by
\begin{equation}
    g_\psi(t) = \frac{-\psi^*(t)}{\sqrt{\int_0^t dt' |\psi(t')|^2}}.
\end{equation}
The Hamiltonian governing the interaction between a single input pulse of radiation with the system and subsequent retrieval of a single output mode is thus given by
\begin{align}
\begin{split} \label{eq:H}
    \hat{H}(t) = &\hat{H}_s(t) + \frac{i}{2}\left[\sqrt{\Gamma}g_\phi(t) \hata^\dagger \hatsigma_- \right. \\
    &+\left. \sqrt{\Gamma} g_\psi^*(t) \hatsigma_+\hatc + g_\phi(t) g_\psi^*(t) \hata^\dagger \hatc - \text{H.c.} \right],
\end{split}
\end{align}
where $\hata$$(\hatc)$ is the annihilation operator associated with the input (output) virtual cavity, and $\hatsigma_-$ is the operator associated with the removal of an excitation in the quantum scatterer governed by Hamiltonian $\hat{H}_s$. This situation is depicted in Fig. \ref{fig:fig1}(a). The detection of a downstream photon is indistinguishably originating from either cavities or the system, and the evolution of the combined system is accompanied by a Lindblad damping term given by
\begin{align} \label{eq:L}
    \hat{L}_0(t) = \sqrt{\Gamma} \hatsigma_- + g_\phi^*(t) \hata + g_\psi^*(t) \hatc.
\end{align}
{\noindent}Further damping terms $\hat{L}_i$ may be readily incorporated, and the time-evolution of the combined system density matrix is found by solution of a master equation of Lindblad form
\begin{equation}    \frac{d\hat{\rho}}{dt} = -i \left[\hat{H}(t), \hat{\rho} \right] + \sum_{i=0}^n D[\hat{L}_i]\hat{\rho},
\end{equation}
where $D[\hat{L}_i]\hat{\rho} = - \frac{1}{2}(\hat{L}_i^\dagger \hat{L}_i \hat{\rho} + \hat{\rho}\hat{L}_i^\dagger \hat{L}_i) + \hat{L}_i\hat{\rho}\hat{L}_i^\dagger$. The master equation is exact (under the usual Born-Markov approximation), but it does not present the most intuitive description of the interaction. The input pulse is represented by a cavity mode that is completely emptied, while the output pulse is treated as the contents of an initially empty cavity mode that  picks up  radiation directly from the input and scattered by the two-level system. In the absence of a scatterer, the wavepacket and its quantum contents would pass  unchanged, [$\psi(t) = \phi(t)$]. That process is governed by the part 
\begin{equation} \label{eq:H0-int}
    \hat{H}_0(t) = \frac{i}{2}\left[g_\phi(t) g_\psi^*(t) \hata^\dagger \hatc - g_\phi^*(t) g_\psi(t) \hatc^\dagger \hata \right].
\end{equation}
of the Hamiltonian \eqref{eq:H}. 
In the following, we shall see, that in the interaction picture with respect to this term, we recover an intuitive picture of a pulse of radiation travelling ``across'' the scatterer while being modified by the interaction. 

%To transform to the appropriate interaction picture, we identify the part of the Hamiltonian responsible for direct transferal between the input and output pulse,
%\begin{equation} \label{eq:H0-int}
%    \hat{H}_0(t) = \frac{i}{2}\left[g_\phi(t) g_\psi^*(t) \hata^\dagger \hatc - g_\phi^*(t) g_\psi(t) \hatc^\dagger \hata \right].
%\end{equation}
%Going to an interaction picture with respect to this part of the Hamiltonian implies solving the cavity-cavity interaction, which means the dynamics will ``follow'' the pulse propagation across the system, as we shall see in the following. We follow the procedure given in \cite{victor_interaction}.

Going to the interaction picture with respect to Eq.~\eqref{eq:H0-int}, entails solving the equations of motion
\begin{align} \label{eq:int-equations_of_motion}
\begin{split}
    \frac{d}{dt}\hata_I(t) &= \frac{1}{2}g_\phi(t) g_\psi^*(t) \hatc_I(t) \\
    \frac{d}{dt}\hatc_I(t) &= -\frac{1}{2}g_\phi^*(t) g_\psi(t) \hata_I(t),
\end{split}
\end{align}
where subscript $I$ denotes the operators in the interaction picture. In the following, we confine ourselves to the case of identical incoming and outgoing modes $\phi(t) = \psi(t)$. The solution to Eq.~\eqref{eq:int-equations_of_motion} is a rotation of the initial operators by angle $\theta(t)$
\begin{align}
\begin{split} \label{eq:int-equations_of_motion-solution}
    \hata_I(t) &= \cos(\theta(t)) \hata_I(0) - \sin(\theta(t)) \hatc_I(0) \\
    \hatc_I(t) &= \cos(\theta(t)) \hatc_I(0) + \sin(\theta(t)) \hata_I(0),
\end{split}
\end{align}
where $\theta(t)$ is defined by
\begin{equation}
    \frac{d\theta(t)}{dt} = - \frac{1}{2} g_\phi^*(t) g_\psi(t),
\end{equation}
and we note that $\theta(t)$ is real due to the choice of identical input and output modes. The solution to this differential equation is
\begin{equation}
    \label{eq:sin2}
    \sin^2\theta(t) \equiv \int_0^t dt' |\phi(t')|^2,
\end{equation}
as can be seen by differentiating both sides with respect to time and noting that with \eqref{eq:sin2} the numerators of $g_\phi(t)$ and $g_\psi(t)$ are exactly $\cos(\theta(t))$ and $\sin(\theta(t))$ respectively. Inserting this solution into the Hamiltonian and Lindblad term gives exactly the expressions applied in the main text
\begin{align}
    \hat{H}_I &= i\sqrt{\Gamma}\phi^*(t)\left[\hat{a}^\dagger\hat{\sigma}_- + \cot2\theta(t)\hat{c}^\dagger\hat{\sigma}_-\right]+\mathrm{H.c.}, \label{eq:H-int}\\
    \hat{L}_I &= \sqrt{\Gamma}\hatsigma_- - 2 \phi(t)\csc 2\theta (t) \hat{c}.\label{eq:L-int}
\end{align}
In this picture, the input pulse mode $\hata$ is lossless (absent from the Lindblad term), but the Hamiltonian contains an important second mode $\hatc$, coupled to $\hata$ through their interaction with the localized scatterer. The combined system in this picture is depicted in Fig.~\ref{fig:fig1}(b). We note that  Eq.~\eqref{eq:int-equations_of_motion-solution} performs a complete rotation from $\theta(t=0) = 0$ to $\theta(t\rightarrow \infty) = \pi / 2$. Hence, the time dependent mode truly reflects the quantum state of the travelling pulse of light. Here, the interaction strength between the scatterer and the time dependent mode $\hata$ is given by the field amplitude of the travelling pulse, similar to the intuitive Jaynes-Cummings (JC) model, but inclusion of the auxiliary mode $\hatc$ can effectively describe the scattering dynamics as well as the exact output state in the input mode.

\begin{figure*}
    \centering
    \includegraphics[width=0.96\linewidth]{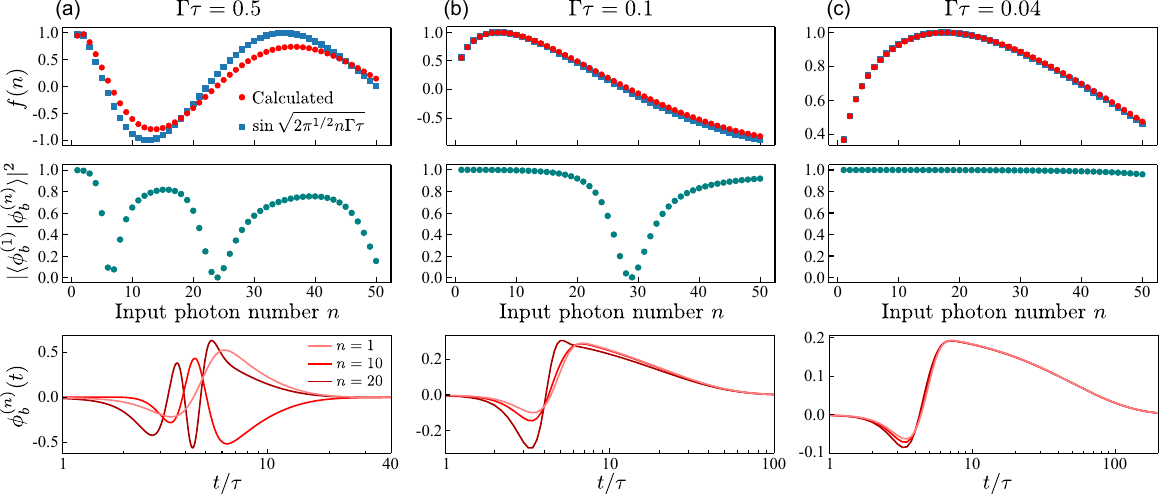}
    \caption{Filtering function $f(n)$, overlap $|\langle \phi_b^{(1)}|\phi_b^{(n)}\rangle|^2$ between the subtracted modes, and mode functions $\phi_b^{(n)}(t)$ for the indicated input photon numbers ($n=1$, $10$, $20$). (a)-(c) show the results for $\Gamma\tau=0.5$, $0.1$, and $0.04$, respectively. In the upper panel, the circles and the squares indicate the results obtained from Eq.~\eqref{eq:s15} and analytical evaluations, respectively.}
    \label{fig:fig2}
\end{figure*}
\section{A General Quantum trajectory formalism}\label{secII}
The master equation derived in Sec.~\ref{sec:SecI}, can be treated very efficiently with the quantum trajectory method. Subtraction of a  single photon  implies a single quantum jump described by the Lindblad operator $\hat{L}_I$. For an $n$-photon input state, the wavefunction for the subtracted photon is thus given by
\begin{equation}
    \tilde{\phi}_b^{(n)}(t) = \langle{n-1,g,0}|{\hat{U}_\mathrm{eff}(\infty,t)\cdot\hat{L}_I\cdot\hat{U}_\mathrm{eff}(t,0)}|n,g,0\rangle,\label{eq:s13}
\end{equation}
where the superscript $(n)$ denotes the $n$-dependence of the wavefunction, and $\hat{U}_\mathrm{eff}(t_2,t_1)$ describes the non-unitary evolution from time $t_1$ to time $t_2$ governed by the effective Hamiltonian $\hat{H}_\mathrm{eff}=\hat{H}_I-i\hat{L}_I^\dagger\hat{L}_I/2$. The norm of the wavefunction $P_n = \int dt\, |\Tilde{\phi}_b^{(n)}(t)|^2$ represents the efficiency of the single-photon subtraction.

We note that Eq.~\eqref{eq:s13} gives the output wavefunction after the pick-up cavity $\hat{c}$ in Fig.~\ref{fig:fig1}(a), which is introduced only for computational purposes. In the experiment, there is no such cavity, and the photon is propagating freely after its emission by the TLE. Taking into account the dispersion induced by the auxiliary mode $\hatc$, the normalized mode function $\phi_b^{(n)}(t)$ for the subtracted single photon right after the TLE is given by 
\begin{equation}
    \phi_b^{(n)} (t) = \frac{\Tilde{\phi}_b^{(n)}(t)}{\sqrt{P_n}}-\phi_a(t)\int_t^\infty d\xi\, \phi_a^*(\xi)\frac{\Tilde{\phi}_b^{(n)}(\xi)}{\sqrt{P_n}}\csc^2\theta(\xi),\nonumber
\end{equation}
which defines the creation operator $\hat{b}_n^\dagger=\int dt \phi_b^{(n)}(t)\hat{\mathcal{E}}^\dagger(t)$ for the mode occupied by the subtracted photon. 

A corresponding no-jump dynamics accounts for the state with all photons remaining in the travelling wave packet and has the state amplitude, $A_n= \langle{n,g,0}|{\hat{U}_\mathrm{eff}(\infty,0)}|n,g,0\rangle$. The output state of the light field is thus given by
\begin{equation}
    \ket{\Psi_\mathrm{out}} = A_n\frac{(\hata^\dagger)^n}{\sqrt{n!}}\ket{0}+\sqrt{P_n}\frac{(\hata^\dagger)^{n-1}}{\sqrt{(n-1)!}}\hatb_n^\dagger\ket{0}+\cdots,
    \label{eq:s14}
\end{equation}
where the ellipsis represents multiphoton subtraction events. While the above discussion holds for any input Fock state, the dependence of mode $\hatb_n$ on the photon number $n$ complicates the calculation of the output state for a general input state of the form $\sum_n c_n|n\rangle$. To get an explicit subtraction operation, we consider a process in which exactly a single photon is converted to mode $\hatb_1$. The input state is then transformed into $\sum_{n\geq1} c_nf(n)|n-1\rangle$ with a filtering function $f(n) = \sqrt{P_n}\langle \phi_b^{(1)}|\phi_b^{(n)}\rangle$, where the overlap
\begin{equation}
    \langle \phi_b^{(1)}|\phi_b^{(n)}\rangle = \int_0^\infty dt\, [\phi_b^{(1)}(t)]^*\phi_b^{(n)}(t)\label{eq:s15}
\end{equation}
quantifies the similarity between the modes. Here, we project all $n$-dependent modes onto $\hatb_1$ mode, because $\phi_b^{(n)}(t)$ usually varies slowly when $n$ is small (see below).

Figure \ref{fig:fig2} illustrates the filtering function as well as the overlap between modes in different regimes. We note that for $\Gamma\tau\sim 1$ [see Fig.~\ref{fig:fig2}(a)], $f(n)$ shows a visible deviation from the analytical expression $\sin\sqrt{2\pi^{1/2}n\Gamma\tau}$ predicted by the intuitive JC model. When $\Gamma\tau$ becomes smaller, $f(n)$ is close to the analytical prediction for a wider range of $n$ [see Fig.~\ref{fig:fig2}(b)]. The mode function $\phi_b^{(n)}(t)$ also varies slowly with the input photon number $n$, except for regions where $f(n)\approx0$. In the regime $\Gamma\tau \ll 1$ [see Fig.~\ref{fig:fig2}(c)], the mode function is almost unchanged, and $f(n)$ approaches the analytical solution.

In the main text, we focus on the regime $n\Gamma\tau<1$, where the truncated Hilbert space illustrated in Fig.~3(c) is sufficient to yield very accurate predictions. With such a truncation, at most one photon can be subtracted, such that Eqs.~\eqref{eq:s13} and \eqref{eq:s14} reduce to Eqs.~(3) and (4) in the main text. In this regime, the continuous field operator can be expanded as $\hat{\mathcal{E}}_\mathrm{out}(t)=\phi_a(t)\hata+\phi_b(t)\hatb$, which facilitates the calculation of the output field correlator
\begin{equation}
    \langle \hat{\mathcal{E}}^\dagger_\mathrm{out}(t)\hat{\mathcal{E}}_\mathrm{out}(t^\prime)\rangle
                  = \begin{bmatrix}
                      \phi_a^*(t) & \phi_b^*(t)
                  \end{bmatrix}
                  \mathbf{G}
                  \begin{bmatrix}
                      \phi_a(t') \\ \phi_b(t')
                  \end{bmatrix},
\end{equation}
where the covariance matrix $\mathbf{G}$ is given by
\begin{equation} 
    \begin{bmatrix}
                     \langle \hata^\dagger\hata \rangle & \langle \hata^\dagger\hatb \rangle \\ \langle \hatb^\dagger\hata \rangle & \langle \hatb^\dagger\hatb \rangle
    \end{bmatrix}  = 
    \begin{bmatrix}
                      n|C_1(\infty)|^2+(n-1)P & C_1(\infty)P\\ C_1^*(\infty)P & P
    \end{bmatrix}.\nonumber
\end{equation}
Exact diagonalization of $\mathbf{G}$ gives the two eigenmodes, whose eigenvalues, $\bar{n}_1$ and $\bar{n}_2$, represent the mean photon number in these modes. In Fig.~2(a) of the main text, we present $\bar{n}_1$ and $\bar{n}_2$ for different pulse durations and find good agreement with exact calculations.

\begin{figure*}
	\centering
	\includegraphics[width=0.96\linewidth]{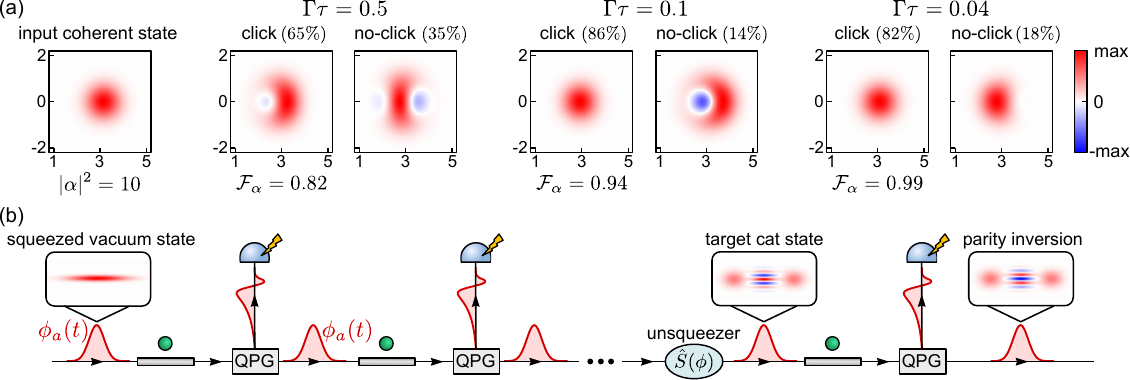}
	\caption{(a) Wigner function of the conditional output states for a coherent input state $|\alpha\rangle=|\sqrt{10}\rangle$ under different interaction strengths $\Gamma\tau$.  The click and no-click events correspond to successful and failed subtractions, respectively, whose probability of occurrence is indicated at the top of each panel. The fidelity $\mathcal{F}_\alpha$ between the photon-subtracted state and the input coherent state is indicated at the bottom of the corresponding panel. (b) Illustration of the setup for generating the Schr\"odinger cat state. The calculations shown here are based on $M=6$, $r=1.15$ (10dB initial squeezing), $\phi=-0.23$, and $\Gamma\tau=0.04$.}
	\label{fig:fig3s}
\end{figure*}

\section{Two-photon scattering}
In addition to the above general quantum-trajectory treatment, we can also understand the subtraction and the mode splitting from a formal scattering theory perspective. Here, we take the two-photon scattering ($n=2$) as an example, which allows us to get an analytical expression for the output state \cite{RevModPhys.95.015002}. In the frequency domain for the input Gaussian two-photon wavefunction $\Psi^{(\text{in})}(\omega_1, \omega_2) = \tau e^{-{(\omega_1^2+\omega_2^2)\tau^2}/{2}} / \sqrt{\pi} $, one can express the output wavefunction as:
\begin{gather}
    \Psi^{(\text{out})}(\omega_1, \omega_2) = \Psi^{(\text{L})}(\omega_1, \omega_2) + \Psi^{(\text{NL})}(\omega_1, \omega_2), \nonumber\\
    \Psi^{(\text{L})}(\omega_1, \omega_2) = \dfrac{\tau e^{-{(\omega_1^2+\omega_2^2)\tau^2}/{2}}}{\sqrt{\pi} } \dfrac{\omega_1 - i \frac{\Gamma}{2}}{\omega_1 + i \frac{\Gamma}{2}} \dfrac{\omega_2 - i \frac{\Gamma}{2}}{\omega_2 + i \frac{\Gamma}{2}},  \nonumber\\
    \Psi^{(\text{NL})}(\omega_1, \omega_2) = \dfrac{\tau\Gamma^2 e^{-{(\omega_1 + \omega_2)^2\tau^2}/{4}} \, w \left( \frac{(\omega_1 + \omega_2 + i \Gamma)\tau}{2} \right)}{\sqrt{\pi}  \left( \omega_1 + i \frac{\Gamma}{2} \right) \left( \omega_2 + i \frac{\Gamma}{2} \right)},
\end{gather}
where $w(z)$ is the Faddeeva function. This is a type of error function, which in the limit $z=z'+iz''$, $z'' \to 0$ ($\Gamma\tau\ll 1$) can be roughly approximated as $\omega(z) \to e^{- z'^2}$, which makes the non-linear part of the wavefunction 
$$\Psi^{(\text{NL})}(\omega_1, \omega_2) \approx \dfrac{\tau\Gamma^2 e^{-{(\omega_1 + \omega_2)^2\tau^2}/{2}} }{\sqrt{\pi}  \left( \omega_1 + i \frac{\Gamma}{2} \right) \left( \omega_2 + i \frac{\Gamma}{2} \right)}.$$
As a further approximation, we can use $e^{-(\omega_1+\omega_2)^2\tau^2/2} \approx e^{-(\omega_1^2+\omega_2^2)\tau^2/2} $, leading to the following approximate wavefunction after the renormalization:
\begin{multline}
    \tilde{\Psi}(\omega_1, \omega_2) \approx \bigg( \dfrac{\tau e^{-(\omega_1^2+\omega_2^2)\tau^2/2}}{\sqrt{\pi} } \dfrac{\omega_1 - i \frac{\Gamma}{2}}{\omega_1 + i \frac{\Gamma}{2}} \dfrac{\omega_2 - i \frac{\Gamma}{2}}{\omega_2 + i \frac{\Gamma}{2}} +  \\
    \dfrac{\Gamma^2\tau e^{-(\omega_1^2+\omega_2^2)\tau^2/2} }{\sqrt{\pi} \left( \omega_1 + i \frac{\Gamma}{2} \right) \left( \omega_2 + i \frac{\Gamma}{2} \right)} \bigg) \times \frac{1}{\sqrt{1 + {2 \pi (\Gamma\tau)^2} w^2\left( i \frac{\Gamma\tau}{2} \right) }}.\nonumber
\end{multline}
The last approximation disentangled the two photons in the non-linear part, allowing us to write the wavefunction in a form that has rank $2$: $\tilde{\Psi}(\omega_1,\omega_2) = c_1 \phi_1(\omega_1) \phi_1(\omega_2) + c_2 \phi_2(\omega_1) \phi_2(\omega_2)$. The normalized (but not orthogonal) functions $\phi_1, \phi_2$ can then be used as a basis to find %representation of the approximate output wavefunction in a form of a $2 \times 2$ matrix, the singular values $\lambda_1, \lambda_2$ of which provide the 
a two-component Schmidt decomposition of the two-photon state coefficients. We will not provide explicit expressions as they are quite cumbersome, but the degeneracy condition for the Schmidt coefficients yields a transcendental  equation for the optimal pulse duration ${2\sqrt{\pi} \Gamma\tau  w\left(i\frac{\Gamma\tau}{2}\right)} = 1$. By solving this equation, one obtains that degeneracy occurs at ${\tau}_\mathrm{opt} \approx 0.34 \Gamma^{-1} $, which is comparable to the true solution for the exact wavefunction which gives $\tau_\mathrm{opt} \approx 0.38\Gamma^{-1}$.

If we have a finite $\beta$ factor, for example, due to emission out of the waveguide or backward scattering, the equation becomes $ {2\sqrt{\pi} \beta \Gamma\tau   w \left( i \frac{\Gamma\tau}{2} \right)} = 1$, which pushes the degeneracy point towards temporally longer pulses, however, for sufficiently large losses and/or backward scattering, this approach might give qualitatively wrong results due to the breakdown of the approximations being made.

\section{Distillation of non-Gaussian states}
To distill useful non-Gaussian states from a Gaussian one, we need to construct the subtraction operation described by the annihilation operator $\hata$. As demonstrated in Sec.~\ref{secII}, the filtering function for a successful subtraction is approximately given by $f(n)\approx\sin\sqrt{2\pi^{1/2}n\Gamma\tau}$ in the regime $\Gamma\tau< 1$, which approaches the ideal one $f(n)\propto\sqrt{n}$ for photon numbers $n\Gamma\tau < 1$.

The above analysis is conditioned on the successful single-photon subtraction, which requires number-resolved detection of a single photon in a specific mode $\hatb_1$. To avoid the use of number-resolved photon detection, we consider the setup shown in Fig.~4(a) of the main text, in which one only needs to detect if there is any photon subtracted from the input mode $\hata$ with a quantum pulse gate (QPG). To describe the conditional output state for such a configuration, we first compute the full density matrix $\hat{\rho}^a(t)$ governed by the master equation evolution $\partial_t{\hat{\rho}}^a=-i [\hat{H}_I, \hat{\rho}^a] + D[\hat{L}_I]\hat{\rho}^a$ with $\hat{\rho}_a(0)=\hat{\rho}_\mathrm{in}$. The no-click output state corresponds to the no-jump evolution of the input state, given by $\hat{\rho}_\mathrm{no-click}=\hat{U}_\mathrm{eff}(\infty,0)\hat{\rho}_\mathrm{in}\hat{U}_\mathrm{eff}^\dagger(\infty,0)$. The photon-subtracted state is then determined by $\hat{\rho}_\mathrm{click}=\hat{\rho}^a(\infty)-\hat{\rho}_\mathrm{no-click}$, and the success probability of the process is $P_s = \mathrm{tr}(\hat{\rho}_\mathrm{click})$.
 
%To demonstrate that our scheme can efficiently and accurately approximate $\hata$, we investigate the conditional output states with a coherent input state $|\alpha\rangle$. As shown in Fig.~\ref{fig:fig3s}(a), a large $\Gamma\tau$ will result in a distorted output state $\hat{\rho}_\mathrm{click}$ having a small fidelity $\mathcal{F}_\alpha$ with the input state. Entering the regime $\Gamma\tau\ll 1$, $\mathcal{F}_\alpha$ becomes close to unity, indicative of a successful implementation of $\hat{a}$, which renders the input state $|\alpha\rangle$ invariant. A larger fidelity can be obtained by further decreasing $\Gamma\tau$, which slightly reduces the success probability. Nonetheless, the efficiency of our scheme is significantly higher than the linear optics scheme at the same operation fidelity, due to the suppression of multiphoton subtractions.

To demonstrate that our scheme can efficiently and accurately approximate the action of  $\hata$, we investigate the conditional output states with a coherent input state $|\alpha\rangle$, which is the eigenstate of $\hata$. As shown in Fig.~\ref{fig:fig3s}(a), rather than leaving the input state $|\alpha\rangle$ invariant, a large $\Gamma\tau\sim 1$ will result in a distorted output state $\hat{\rho}_\mathrm{click}$ with a small fidelity $\mathcal{F}_\alpha=\langle\alpha|\hat{\rho}_\mathrm{click}|\alpha\rangle/P_s$. Notably, a very high fidelity can be obtained by only moderately reducing $\Gamma\tau$, at which the success probability is still considerable. Therefore, our protocol can herald high fidelity states with much higher success probabilities than the linear optics schemes that are severely limited by the multiphoton subtraction events.

By mimicking $\hata$, one can generate useful non-Gaussian states with high fidelities. As shown in Fig.~4 of the main text, we consider generating a Schr\"odinger cat state $|\mathrm{cat}\rangle\propto [|\alpha\rangle+(-1)^M|-\alpha\rangle]$ with $\alpha=\sqrt{M}$ from a squeezed vacuum state $\hat{S}(r)|0\rangle$ with $\hat{S}(r)=e^{-r(\hat{a}^2-\hat{a}^{\dagger2})/2}$. For a sufficiently large initial squeezing ($r>1$), after $M$ successive photon subtractions, the input state approaches a squeezed cat state, which after an optimal anti-squeezing process $\hat{S}(\phi)$, will reach the target state $|\mathrm{cat}\rangle$, as illustrated in Fig.~\ref{fig:fig3s}(b). The parity of the generated cat state can be inverted by another subtraction operation.

\section{Proof of the reciprocity}
In this section, we derive a reciprocity relation between the single-photon subtraction and addition processes. To this end, we first consider the total Hamiltonian of the chiral waveguide QED system in the frame rotating with the transition frequency $\omega$ of the TLE (not to be confused with the moving frame) \cite{RevModPhys.89.021001}
\begin{align}
    \hat{H}=\int_{-\infty}^\infty dk \, v_gk\hat{a}_k^\dagger\hat{a}_k+ \sqrt{\frac{v_g\Gamma}{2\pi}}\int_{-\infty}^\infty dk \, (\hat{a}_k^\dagger\hat{\sigma}_- + \hat{\sigma}_+\hat{a}_k),\label{eq:s19}
\end{align}
where $v_g$ denotes the (assumed constant) group velocity of the photons. $\hat{a}_k^\dagger$ creates a photon with an additional momentum $k$ on top of the center momentum $k_0=\omega/v_g$, and obeys the commutation relation $[\hat{a}_k,\hat{a}_{k^\prime}^\dagger]=\delta({k-k^\prime})$. The field operator evaluated at $x=0$ (the location of the TLE) is given by $\hat{\mathcal{E}}(t)=\int_{-\infty}^\infty dk  e^{-iv_gkt}\hat{a}_k/\sqrt{2\pi}$, which satisfies the commutation relation $[\hat{\mathcal{E}}(t),\hat{\mathcal{E}}^\dagger(t^\prime)]=\delta({t-t^\prime})$. Introducing the parity operator $\mathcal{P}$ and the time reversal operator $\mathcal{T}$:
\begin{align}
    \mathcal{P} \hat{a}_k \mathcal{P}^{-1} &= \hat{a}_{-k},\
    \mathcal{P} \hat{\mathcal{E}}(t) \mathcal{P}^{-1}= \hat{\mathcal{E}}(-t),\
    \mathcal{P} i \mathcal{P}^{-1}= i,\nonumber\\
    \mathcal{T} \hat{a}_k \mathcal{T}^{-1} &= \hat{a}_{-k},\
    \mathcal{T} \hat{\mathcal{E}}(t) \mathcal{T}^{-1}= \hat{\mathcal{E}}(t),\
    \mathcal{T} i \mathcal{T}^{-1}= -i,\nonumber
\end{align}
we find that the system is invariant under the combined action of $\mathcal{P}$ and $\mathcal{T}$, i.e., $\mathcal{P}\mathcal{T}\hat{H}(\mathcal{P}\mathcal{T})^{-1}=\hat{H}$. The parity-time symmetry of the system then endows the unitary evolution operator $\hat{U}(t_f,t_i)$ with the property
\begin{equation}
    \mathcal{P}\mathcal{T}\hat{U}(t_f,t_i)(\mathcal{P}\mathcal{T})^{-1}=\hat{U}(t_f,t_i)^\dagger,\label{eq:s20}
\end{equation}
which leads to the following relation
\begin{align}
\nonumber
    |\mel{\mathrm{out}}{\hat{U}(T,0)}{\mathrm{in}}|^2
& =  |\mel{\mathrm{in}}{\hat{U}(T,0)^\dagger}{\mathrm{out}}|^2\\
& =  |\mel{\mathrm{in}}{ \mathcal{P}\mathcal{T}\hat{U}(T,0)(\mathcal{P}\mathcal{T})^{-1}}{\mathrm{out}}|^2.\nonumber
\end{align}
With $(\mathcal{P}\mathcal{T})^2=1$ and $\mathcal{P}\mathcal{T}(\mathcal{P}\mathcal{T})^\dagger=-1$, we finally arrive at the reciprocity theorem
\begin{equation}
    |\mel{\mathrm{out}}{\hat{U}(T,0)}{\mathrm{in}}|^2 =  |\mel{\mathrm{in}}{ (\mathcal{P}\mathcal{T})^\dagger\hat{U}(T,0)\mathcal{P}\mathcal{T}}{\mathrm{out}}|^2.\label{eq:s21}
\end{equation}
Eq.~\eqref{eq:s21} implies that the overlap between the target state $|\mathrm{out}\rangle$ and the time-evolved state $\hat{U}(T,0)|\mathrm{in}\rangle$ is identical to the overlap between the target state $\mathcal{PT}|\mathrm{in}\rangle$ and the time-evolved state $\hat{U}(T,0)\mathcal{PT}|\mathrm{out}\rangle$.

Next, we apply the above reciprocity theorem to the photon subtraction/addition. We will only consider temporal modes $\hata$ and $\hatb$ described in the main text. Here, we notice that $\hata$ and $\hatb$ are defined in the moving frame, while Hamiltonian Eq.~\eqref{eq:s19} and the above derivation work for the static rotating frame, where the creation operator $\hata^\dagger$ and $\hatb^\dagger$ become time dependent
\begin{align}
    \hat{A}^\dagger(t)&=\int_{-t}^{T-t} d\xi\ \phi_a(\xi+t)\hat{\mathcal{E}}^\dagger(\xi), \nonumber \\ \hat{B}^\dagger(t)&=\int_{-t}^{T-t} d\xi\ \phi_b(\xi+t)\hat{\mathcal{E}}^\dagger(\xi).\nonumber
\end{align}
For the single-photon subtraction, the input state is
\begin{equation}
|\mathrm{input}\rangle=|n,0\rangle={[\hat{A}^\dagger(0)]^n}|0\rangle / {\sqrt{n}},
\end{equation}
the target output state is
\begin{equation}
|\mathrm{output}\rangle=|n-1,1\rangle=\frac{[\hat{A}^\dagger(T)]^{n-1}}{\sqrt{n-1}}\hat{B}^\dagger(T)|0\rangle,\label{eq:s23}
\end{equation}
and the success probability is given by
\begin{equation}
    P_{n}^\mathrm{s} = |\mel{n-1,1}{\hat{U}(T,0)}{n,0}|^2. \label{eq:s24}
\end{equation}
Then, by choosing $\mathcal{PT}|n-1,1\rangle$ as the input state and $\mathcal{PT}|n,0\rangle$ as the output state, we construct a single-photon addition process, whose success probability reads
\begin{equation}
    P_{n}^\mathrm{a} = |\mel{n,0}{(\mathcal{PT})^\dagger\hat{U}(T,0)\mathcal{PT}}{n-1,1}|^2.\label{eq:s25}
\end{equation}
With the reciprocity theorem Eq.~\eqref{eq:s21}, we prove that $P_{n}^\mathrm{a} =P_{n}^\mathrm{s}$. Here, the parity-time operator transforms output modes in Eq.~\eqref{eq:s23} into input modes:
\begin{align}
\nonumber
\hat{A}_T^{\dagger}(0)=\mathcal{PT}\hat{A}^\dagger(T)(\mathcal{PT})^{-1}&=
\int_{0}^{T} d\xi\ \phi_a^*(T-\xi)\hat{\mathcal{E}}^\dagger(\xi),\\
\nonumber
\hat{B}_T^{\dagger}(0)=\mathcal{PT}\hat{B}^\dagger(T)(\mathcal{PT})^{-1}&=
\int_{0}^{T} d\xi\ \phi_b^*(T-\xi)\hat{\mathcal{E}}^\dagger(\xi).
\end{align}
The above relation yields the optimal input mode for the single-photon addition, i.e., by choosing temporal modes $\phi_a^T(t)=\phi_a^*(T-t)$ and $\phi_b^T(t)=\phi_b^*(T-t)$ in the moving frame, which carry $n-1$ photons and a single photon, respectively, we can generate an $n$-photon Fock state in mode $\phi_a^T(t)$ with a success probability $P_n^\mathrm{a}$.

The reciprocity relation $P_{n}^\mathrm{a} =P_{n}^\mathrm{s}$ is verified by numerical simulations shown in Fig.~5(a) of the main text. For the subtraction process, we use one virtual input cavity to emit photons in mode $\phi_a(t)$ and one virtual output cavity to catch photons in mode $\phi_a(t)$. The success probability $P_{n}^\mathrm{s}=\rho_{n-1,n-1}^a=\langle n-1|\hat{\rho}^{a}|n-1\rangle$ is extracted from the reduced density matrix $\hat{\rho}^{a}$ of the output virtual cavity, which is equivalent to the definition Eq.~\eqref{eq:s24}. For the addition process, we use two virtual input cavities to emit photons in mode $\phi_a^T(t)=\phi_a^*(T-t)$ and $\phi_b^T(t)=\phi_b^*(T-t)$, and a single virtual output cavity to catch photons in mode $\phi_a^T(t)=\phi_a^*(T-t)$. Here, to improve the accuracy of the calculation, we use the general formalism Eq.~\eqref{eq:s13} to obtain the subtracted mode function $\phi_b(t)$. The success probability $P_{n}^\mathrm{a}=\rho_{n,n}^a=\langle n|\hat{\rho}^{a}|n\rangle$ is extracted from the reduced density matrix $\hat{\rho}^{a}$ of the output virtual cavity, equivalent to the definition Eq.~\eqref{eq:s25}.

\section{Experimental Considerations}
\subsection{Tunable two-level nonlinearity}
As discussed in the main text, in order to achieve near-deterministic subtraction or addition, the input pulse duration needs to be close to the optimal value $\tau_\mathrm{sub}$, which depends on both the atom-photon interaction strength and the input photon number $n$, i.e., $\tau_\mathrm{sub}\sim 1/(\Gamma n)$. While the pulse may be reshaped via some active schemes, e.g., the quantum pulse gate \cite{eckstein2011quantum}, it is equivalent to fix the pulse duration $\tau$ and tune the decay rate $\Gamma$ of the TLE. Here, we list several realistic schemes for achieving a tunable $\Gamma$.

In a quantum photonic platform \cite{lodahl2}, the atom-photon interaction strength $G \propto \sqrt{v_g \Gamma}$ can be engineered by adjusting the position of the TLE and designing the field distribution of the waveguide mode. We can also tune the two-level nonlinearity in a $\Lambda$-type, three-level emitter (3LE) involving two stable states $\ket{g},\ket{s}$ and an intermediate excited state $\ket{e}$, where the $|g\rangle-|e\rangle$ transition is driven by the quantum field $\hat{\mathcal{E}}(t)$ with an interaction strength $G$, while the $|e\rangle-|s\rangle$ transition is driven by a classical coherent field with a Rabi frequency $\Omega$ \cite{PhysRevResearch.4.023002}. By choosing a large detuning $\Delta$ from the intermediate state, one can adiabatically eliminate state $|e\rangle$ and obtain an effective TLE interacting with the quantum probe field $\hat{\mathcal{E}}(t)$ with a tunable coupling strength $G_\mathrm{eff}\propto G\Omega/\Delta$.

\begin{figure}
	\centering
	\includegraphics[width=\linewidth]{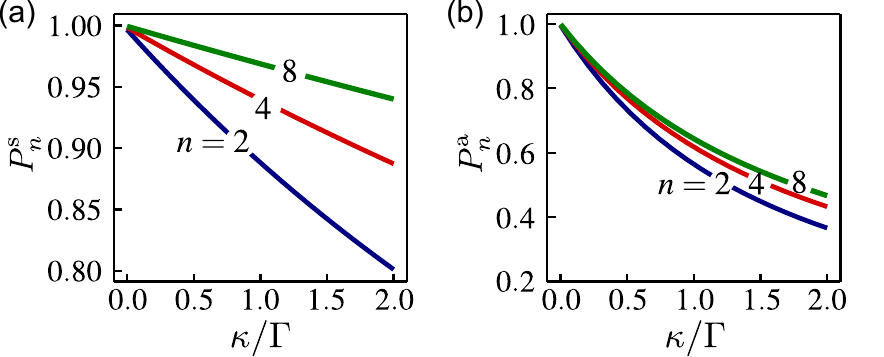}
	\caption{(b) and (c) show the single-photon subtraction and addition success probability with a finite dephasing rate $\kappa$.}
	\label{fig:fig4s}
\end{figure}
In circuit-QED platforms, superconducting circuit components offer a large degree of engineering, making it possible to achieve strong and tunable coupling between elements. For example, a transmon qubit can serve as a TLE and be capacitively coupled to a transmission line supporting a 1D continuum microwave field \cite{circuitQED}. By varying the capacitive coupling, the TLE acquires a tunable spontaneous decay rate $\Gamma$ to the transmission line.

The desired nonlinearity can also be achieved by using Rydberg superatoms in free space \cite{stiesdal2021} or in a cavity \cite{PhysRevX.12.021034}.

\subsection{Imperfection of the TLE}
In a realistic setup, the coupling between the TLE and the unidirectional waveguide mode of interest is not always perfect. To describe the coupling between the TLE and the unwanted photonic modes, we include an additional Lindblad term $\hat{L}_\gamma = \sqrt{\gamma} \hatsigma_-$ in the effective master equation, where $\gamma$ quantifies the incoherent decay rate. The dependence of success probabilities for the single-photon subtraction ($P_n^\mathrm{s}$) and addition ($P_n^\mathrm{a}$) on $\gamma$ are shown in Figs.~5(b) and 5(c) of the main text.

In addition to the decay term, the coupling between the
TLE and the environment can also cause a pure dephasing, which can be described by an individual Lindblad dissipation term $\hat{L}_\kappa=\sqrt{\kappa}\hat{\sigma}_{ee}$. Figures \ref{fig:fig4s}(a) and \ref{fig:fig4s}(b) display the success probabilities $P_n^\mathrm{s}$ and $P_n^\mathrm{a}$ as function of the dephasing rate $\kappa$. Similar to the case with an imperfect coupling, $P_n^\mathrm{s}$ is much more robust against dephasing than $P_n^\mathrm{s}$, because the photon subtraction process is more insensitive to the multimode character of the subtracted single photon.

\bibliography{main_text_bib}